\documentclass[amsmath,amssymb,showpacs,showkeywords]{revtex4}
\usepackage{amsmath,amsfonts,amssymb,latexsym,extarrows}
\newtheorem{thm}{Theorem}
\newtheorem{lemma}{Lemma}

\newtheorem{rema}{Remark}

\renewcommand{\theequation}{\arabic{equation}}

\newcommand{\dis}{\displaystyle}

\newcommand{\bequ}{\begin{equation}}
\newcommand{\eequ}{\end{equation}}
\newcommand{\barr}{\begin{array}}
\newcommand{\earr}{\end{array}}
\newcommand{\bea}{\begin {eqnarray}}
\newcommand{\eea}{\end {eqnarray}}
\newcommand{\lb}{\label}

\newcommand{\qed}{\hfill \rule{2.25mm}{2.25mm}\vspace{.15cm}}

\renewcommand{\Im}{{\cal I}{\rm m}\:}
\renewcommand{\Re}{{\cal R}{\rm e}\:}

\begin{document}
\def \tr {\mathrm{Tr\,}}
\let\la=\lambda
\def \Z {\mathbb{Z}}
\def \Zt {\mathbb{Z}_o^4}
\def \R {\mathbb{R}}
\def \C {\mathbb{C}}
\def \La {\Lambda}
\def \ka {\kappa}
\def \vphi {\varphi}
\def \Zd {\Z ^d}
\title{On Yang-Mills Stability and Plaquette Field Generating Functional}
\author{Michael O'Carroll}\email{michaelocarroll@gmail.com}
\author{Paulo A. Faria da Veiga}\email{veiga@icmc.usp.br. Orcid_Registration _Number: 0000-0003-0739-069X.}
\affiliation{Departamento de Matem\'atica Aplicada e Estat\'{\i}stica - ICMC, USP-S\~ao Carlos,\\C.P. 668, 13560-970 S\~ao Carlos SP, Brazil}
\date{May 02, 2020.} 
\begin{abstract}
 We consider the pure Yang-Mills relativistic quantum field theory in an imaginary time functional integral formulation.  The gauge group is taken to be $\mathcal G = \mathrm U(N)$. We use a lattice ultraviolet regularization, starting with the model defined on a finite hypercubic lattice $\Lambda\subset a\mathbb Z^d$, $d = 2,3,4$, with lattice spacing $a\in (0,1]$ and $L\in\mathbb N$ sites on a side.  The Wilson partition function is used where the action is a sum over four lattice bond variables of gauge-invariant plaquette (lattice minimal squares) actions with a prefactor $a^{d-4}/g^2$, where we take the gauge coupling $g\in(0,g_0^2]$, $0<g_0<\infty$. In a recent paper, for free boundary conditions, we proved that a normalized model partition function satisfies thermodynamic and ultraviolet stable stability bounds. Here, we extend the stability bounds to the Yang-Mills model with periodic boundary conditions, with constants which are also independent of $L$, $a$, $g$. Furthermore, we also consider a normalized generating functional for the correlations of $r\in\mathbb N$ gauge-invariant plaquette fields. Using periodic boundary conditions and the multireflection method, we then prove that this generating functional is bounded, with a bound that is independent of $L$, $a$, $g$ and the location and orientation of the $r$ plaquette fields. The bounds factorize and each factor is a single-bond variable, single-plaquette partition function. The number of factors is, up to boundary corrections, the number of non-temporal lattice bonds, such as $(d-1)L^d$. A new global quadratic upper bound in the gluon fields is proved for the Wilson plaquette action. 
\end{abstract}
\maketitle
%
\section{Introduction} \lb{intro}
To show the existence and properties of an interacting relativistic quantum field theory (QFT) in spacetime dimension four is a fundamental problem in physics \cite{Wei,Banks,Gat,GJ}. Many partial results have been obtained \cite{GJ,Riv,Summers,Sei}.  The quantum chromodynamics model (QCD) of interacting (anti)quarks and gauge, gluon fields is considered to be a good candidate for a four dimensional QFT model which rigorously exists. The action of this model is a sum of an interacting Fermi-gauge field part and a pure-gauge field self-interacting Yang-Mills (YM) part.

In this paper, we will focus only on the pure-gauge YM model. In an imaginary-time functional integral formulation, a lattice ultraviolet regularization is used.  The starting point is the Wilson plaquette action partition function.  Stability bounds (see \cite{Rue}) for the corresponding partition function have been proved in the seminal  work of Balaban (see \cite{Bal,Bal2} and Refs. therein), using renormalizationg group methods and the heavy machinery of multiscale analysis. Using softer methods, in Ref. \cite{Ash}, the $d=2$ YM model was shown to exist.  It is expected that partition function stability bounds lead to bounds on field correlations. Indeed, in the context of the renormalization group, considering models which are small perturbations of the free field, the generating functional of field correlations and the correlations can be obtained through a formula which involves the effective actions generated applying the renormalization group transformations to the partition function (see e.g. \cite{MB}). However, in the case of gauge fields, this question, as well as the incorporation of fermion fields, have never been analyzed up to now.

Recently, in \cite{YM}, a simple proof of thermodynamic and ultraviolet stable (TUV) stability bounds is given by a direct analysis of the Wilson partition function with free boundary conditions in configuration space, starting with the model in a finite hypercubic lattice.  The gauge group is taken as $\mathcal G = \mathrm U(N)$ or $\mathrm{SU}(N)$.  For each lattice bond there is a bond variable $U$ which is an element of the gauge group $\mathcal G$. Also, by the spectral theorem, as $U$ is unitary, there exists a unitary $V$ which diagonalizes $U$, i.e. $V^{-1}U V=\mathrm{diag} (e^{i\lambda_1},\ldots,e^{i\lambda_N})$,  $\lambda_j\in(-\pi,\pi]$.  The $\lambda_j$ are called the angular eigenvalues of $U$. The action is a sum over products of four bond variables corresponding to the gauge-invariant Wilson plaquette actions, and with a prefactor $a^{d-4}/g^2$, where we take $g^2\in(0,g_0^2]$, $0<g_0<\infty$.  The hypercubic lattice $\Lambda\subset a\mathbb Z^d$, $d=2,3,4$, with spacing $a\in(0,1]$ has $L\in\mathbb N$ sites on each side. In \cite{YM}, the finite lattice partition function with free boundary conditions (b.c.) is denoted by $Z_{\Lambda,a}$. Looking foward to considering a more general b.c., in this paper, the partition function will be denoted by $Z^B_{\Lambda,a}$, where  $B$ is left blank, for free b.c., and $B=P$, for periodic b.c. A complete description of the model is given in section \ref{sec2}.

Associated with these classical statistical mechanical model partition functions and its gauge-invariant correlations, there is a a lattice quantum field theory.  The Osterwalder-Seiler \cite{Sei} construction provides, via a Feynman-Kac formula, a quantum mechanical Hilbert space, self-adjoint mutually commuting spatial momentum operators and a positive energy operator. A key property in the construction is Osterwalder-Seiler reflection positivity (ensured here if $L$ is chosen to be even).

It is to be emphasized that the work of Ref. \cite{YM} concentrated only on the existence of a finite normalized free energy for the model, in the (subsequential) thermodynamic and continuum limits, respectively, $\Lambda\nearrow a\mathbb Z^d$ and $a\searrow 0$. No other property of the model was considered.
It is also worth noticing that the techniques and methods used in Ref. \cite{YM}, combined with the results of Refs. \cite{M,MP,MP2} could be combined and used to prove the existence of a normalized free energy for a bosonic lattice QCD model, with the (anti)quark fields replaced with spin zero, multicomponent complex or real scalar fields. This is the content of Ref. \cite{bQCD}.

In this paper, first, we extend the TUV stability bounds to the YM model with periodic b.c. The upper and lower stability bounds have an interesting structure. They are both products of single-plaquette, single-bond variable partition functions.  A new, global quadratic upper bound in the gluon fields for the Wilson plaquette action is proved. This bound gives rise to the factorized lower bound on the partition function. We denote by $z_u$ ($z_\ell$) the single-bond Haar integral partition functions describing  the single-plaquette partition function for the upper (lower) stability bound on the partition function with periodic b.c.  The integrands of $z_u$ and $z_\ell$ are both class functions of the single variable $U$, where we recall that a class function $f(U)$ on the gauge group $\mathcal G$ satisfies the property $f(U)= f(VUV^{-1})$ for all $V\in\mathcal G$. Thus, by Weyl’s integration formula \cite{Weyl,Bump,Simon2} the $N^2$-dimensional (for $\mathcal G=\mathrm{U}(N)$ Haar integration over the group is reduced to an $N$-dimensional integration over the angular eigenvalues of $U$. The probability density of the circular unitary ensemble(CUE) occurs and in the bounds on $z_u$ and $z_\ell$ the probability density for the Gausssian unitary ensemble (GUE)  of random matrix theory appear in a natural way (see \cite{Metha,Deift}). The new stability bounds are the contents of Theorem \ref{thm1} below.

Furthermore, here, we also consider a normalized generating functional for the correlation of $r\in\mathbb N$ gauge-invariant plaquette fields \cite{Schor}. The numerator is the periodic b.c. partition function with $r$ additional source factors of strengths $J_j$, $j = 1,\ldots,r$;  the denominator is the periodic b.c. partition function $Z^P_{\Lambda,a}$. Starting with the model with periodic boundary conditions which allows us to apply the multireflection method \cite{GJ}, in Theorem \ref{thm2} below, we prove that this normalized generating functional is absolutely bounded, with a bound that is independent of $L$, $a$, $g$, and the location and orientation of the $r$ {\em external} plaquette fields. The generating functional bound also has an interesting structure. The bound has only a product of single-plaquette, single bond-variable partition function $z_u(J)$ with a source strength field $J$ in the numerator;  in the denominator only a product of $z_\ell$ (the same as in the preceding case!) occurs. In the bound for $z_u(J)$  the probability density for the Gausssian symplectic ensemble appears (see \cite{Deift}). The generating functional is jointly analytic, entire function in the source strengths $J_1$, ..., $J_r$ of the $r$ plaquette fields. The $r$-plaquette field correlations admit a Cauchy integral representation and are bounded by Cauchy bounds. In particular, the coincident point plaquette-plaquette physical field correlation is bounded by $\mathrm{const}\,a^{2-d}$. The $a^{2-d}$ factor, small $a$ behavior is the same as that of the unscaled real scalar free field two-point correlation (see Appendix A).

As the methods of analysis and TUV stability bound results for the free b.c. model play an important role in our treatment of the periodic model, we review the free b.c. case in Section \ref{sec2}, as well as  give the TUV stability bounds for the periodic b.c. model. Our stability bound results are established in Theorems 1-3 below. The normalized generating functional for plaquette field correlations is defined in section \ref{sec3} and its boundedness is established in Theorem 4.  Section \ref{sec4} is devoted to prove the Lemma \ref{lema1}, which gives an upper bound on the Wilson plaquette action, and all the theorems and some concluding remarks are made in \ref{sec5}. Finally, in Appendix A, considering the case of the real scalar free field $\phi$, we develop the relation between quantities expressed in terms of the unscaled or physical field $\phi^u(x)$ and locally scaled fields $\phi(x)\,=\,s(a)\,\phi^u(x)$.
\section{Thermodynmic and Ultraviolet Stability Bounds}\lb{sec2}
We first describe the free and periodic b.c. models and then we give the factorized TUV stability bounds. The superscript $P$ will denote periodic b.c. quantities.  For the lattice $\Lambda$,  we denote by $\Lambda_s=L^d$  the total number of lattice sites.  We let $x =(x^0,\ldots,x^{d-1})$ denote a site, and $x^0$ is the time direction.

\noindent Free b.c. Bonds: Let $e^\mu$, $\mu=0,1,\ldots,(d-1)$ denote the unit vector in the $\mu$-th direction. $b_\mu(x)$ is the lattice bond with initial point $x$ and terminal point $x_+^\mu\equiv x + ae^\mu \in\Lambda$. The number of free b.c. bonds in the lattice $\Lambda$ is $\Lambda_b=d(L-1)L^{d-1}$. Sometimes, we refer to the bonds in the time direction $x^0$ as {\em vertical} bonds. The other bonds are called {\em horizontal}.

\noindent Periodic b.c. Bonds:  In addition to the above free bonds, here, we have additional or extra bonds.  An extra bond has initial point at the extreme right lattice site and terminal point at the extreme left lattice site, in each coordinate direction.  If $\Lambda_e$ denotes the number of extra bonds, we have $\Lambda_e=dL^{d-1}$.  The total number bonds in $\Lambda$ with periodic b.c. (henceforth called {\em periodic bonds}) is $\Lambda_b^P=\Lambda_b+\Lambda_e$.

\noindent Free b.c. Plaquettes (Minimal Lattice Squares): For $\mu,\nu=0,\ldots,(d-1)$, let $p_{\mu\nu}(x)$ denote a plaquette in the $\mu\nu$-plane, with $\mu<\nu$ with and vertices at sites $x$, $x + ae^\mu$, $x + ae^\mu+ae^\nu$, $x + ae^\nu$ of $\Lambda$.  These are the free plaquettes.

\noindent Periodic b.c. Plaquettes:  In addition to the free plaquettes, there are also extra plaquettes formed at least with one extra bond. The periodic b.c. plaquettes are comprised of all plaquettes that can be formed from the totality of periodic b.c. bonds. We denote the total number of free (periodic) plaquettes by $\Lambda_p$ ($\Lambda_p^P$).

Recalling that $a\in(0,1]$ and $g^2\in(0,g^2_0)$, $0<g^2_0,\infty$, we represent the model partition function, with $B$-type b.c., by
\bequ\lb{part}
Z^B_{\Lambda,a}\,=\,\int\, \exp\left[-\,\dfrac{a^{d-4}}{g^2}\,A^B\right]\,dg^B\,.
\eequ
Here, for each lattice bond $b$, we assigned a unitary matrix $U\in U(N)$.  These are the gauge bond variables.  The measure $dg^B$ is the product over bonds $b$ of the single-bond gauge group Haar measures $d\sigma(U)$. For $p$ denoting any fixed plaquette, the model action is given by
\bequ\lb{action}A^B= \sum_p  \,A_p\,,\eequ
where the four bond variable plaquette actions $A_p$ and where the sum $\sum_p$ is over plaquettes with the b.c. of type $B$.

To define $A_p$. we first recall some important facts about unitary matrices and their representation in terms of elements of the Lie algebra of self-adjoint matrices associated with the gauge group $\mathcal G$.  For an $N\times N$ matrix $M$ the Hilbert-Schmidt norm is $\|M\|_{H-S}= [Tr(M^\dagger M)]^{1/2}$, where $M^\dagger$      is the adjoint of $M$.  Let $M_1$ and $M_2$ be $N\times N$ matrices.  Then $(M_1,M_2)=Tr(M_1^\dagger M_2)$ is a sesquilinear inner product.  We also have the following properties:
\begin{enumerate}
\item Let $X$ be a self-adjoint matrix. Define $\exp(iX)$ by the Taylor series expansion of the exponential.  Then $\exp(iX)$ is unitary.
\item Given a unitary $N\times N$ matrix $U$, by the spectral theorem, there exists a unitary $V$ such that $V^{-1}UV= \mathrm{diag}(e^{i\lambda_1}, \ldots,e^{i\lambda_N})$, $\lambda_j\in(-\pi,\pi]$. The $\lambda_j$ are the angular eigenvalues of $U$.  Define $X=V^{-1}\mathrm{diag}(\lambda_1,\ldots,\lambda_N)V$. Then $U= \exp(iX)$, and the exponential map is onto (see \cite{Bump}).
\item For $\alpha= 1,2,\ldots,N$, let the self-adjoint $\theta_\alpha$ form a basis for the self-adjoint matrices (the Lie algebra $u(N)$ generators), with the normalization condition $\mathrm{Tr} \theta_\alpha \theta_\beta=\delta_{\alpha\beta}$, with a Kronecker delta.  Then, with $X$ being an $N\times N$ self-adjoint matrix, $X$ has the representation $X=\sum_{1\leq \alpha\leq N^2} \;x_\alpha\theta_\alpha= \mathrm{Tr} X\theta$, with $x_\alpha\,=\,\tr X\theta_\alpha$.
\item For $U$ and $X$ related as in item $2$, we have the important inequality:
$$
\| X\|^2_{H-S}\,=\,\mathrm{Tr}\left(X^\dagger X\right)\,=\,\sum_{1\leq\alpha\leq N^2}|x_\alpha|^2=|x|^2\,=\,\sum_{1\leq  j\leq N}\,\lambda_j^2\leq N\pi^2 \quad ,\quad \,\lambda_j\in(-\pi,\pi]\,.
$$
Thus, the exponential map is onto, for $|x|\leq N^{1/2}\pi$.
\end{enumerate}

For each bond $b$, we assign the gauge bond variable $U\in\mathrm U(N)$.  If we parametrize $U$ as $e^{iagA_b}$, with $A_b$ self-adjoint, we call $A_b$ the physical gluon field associated with bond $b$. The physical gluon field $A_b$ has the representation $A_b=\sum_{1\leq  j\leq N^2} A_{b}^\alpha \theta_\alpha$, and we refer to $A_{b}^\alpha$, $\alpha=1,\ldots,N^2$, as the color or gauge components of $A_b$. If the plaquette $p$ is $p_{\mu\nu}(x)$, define 
$$U_p\,=\, e^{iagA_\mu(x)}\,e^{iagA_\nu(x+ae_\mu)}\,e^{-iagA_\mu(x+ae_\nu)}\,e^{-iagA_\nu(x)}\,.$$
The plaquette action $A_p$ for the plaquette $p$ is defined by 
\bequ\lb{Ap}
A_p\,=\,\|U_p-1\|^2_{H-S}\,=\,2\,\Re\tr(1-U_p)\,=\,2\tr(1–\cos X_p)\,,
\eequ
where $U_p= e^{iX_p}$. Obviously, $A_p$ is pointwise positive (nonnegative) and so is the total action for the model $A^B=\sum_p\,A_p$.

This completes the description of the model. Using the Baker-Campbell-Hausdorff formula, formally, it is shown in Ref. \cite{Gat}, for small lattice spacing $a>0$, 
$$U_p\,=\,\exp\left[ia^2g\,F^a_{\mu\nu}(x)\,+\,R\right]\quad,\quad  R\,=\,\mathcal O(a^3)\,,$$
where $F_{\mu\nu}^a(x)\,=\,\partial^a_\mu A_\nu(x)\,–\,\partial^a_\nu A_\mu(x)\,+\,ig[A_\mu(x),A_\nu(x)]$ is the usual color 'electromagnetic' second order antisymmetric tensor with finite difference derivatives $\partial^a_\mu A_\nu(x)\,=\,a^{-1}\,\{A_\nu(x+ae_\mu)–A_\nu(x)\}$, and $[\cdot ,\cdot ]$ denotes the Lie algebra commutator (Lie product) associated with the gauge group $\mathcal G$.  Also it is shown that
$A_p\,\simeq\,a^4g^2\,\tr\left[F^a_{\mu\nu}(x)\right]^2$.

Each term in $\left[F^a_{\mu\nu}(x)\right]$ is self-adjoint. Hence, the square is self-adjoint and positive, as well as its trace.  The quantity $[a^{d-4}/g^2]\,\sum_p\,A_p$ is the Riemann sum approximation to the classical smooth field continuum YM action $\int\,\,\tr(F^a_{\mu\nu})^2(x)\,d^dx$, when $\Lambda\nearrow \mathbb Z^d$ and $a\searrow 0$, formally, and the finite difference derivatives become ordinary partial derivatives.

We now discuss gauge invariance and gauge fixing. Due to the local gauge invariance of the action $A_p$, and so also $A=\sum_p\,A_p$, there is an excess of gauge variables in the definition of the partition function given by Eq. (\ref{part}). By a gauge fixing procedure we eliminate gauge variables by setting them equal to the identity in the action and dropping the gauge bond variable integration.  In this process of gauging away some of the gauge group, bond variables, the value of the partition function is unchanged, as long as we apply this procedure to bonds which do not form a closed loop in $\Lambda$ (see \cite{GJ}).  We will use what we call the enhanced temporal gauge.

In the enhanced temporal gauge, the temporal bond variables in $\Lambda$  are set to the identity, as well as certain specified bond variables on the boundary $\partial\Lambda$ of $\Lambda$.  Letting $\Lambda_r$ denote the number of retained bonds, for free b.c., we have $\Lambda_r=(L-1)^2$, $[(2L+1)(L-1)^2]$, $[(3L^3-L^2-L-1)(L-1)]$, respectively, for
$d=2,3,4$. Clearly $\Lambda_r\simeq (d-1) L^d$, for sufficiently large $L$, and $\Lambda_r\nearrow\infty$ as $\Lambda\nearrow a \mathbb Z^d$. For periodic b.c., the same bond variables are gauged away; the number of non-gauged away bond variables is then $\Lambda_r+\Lambda_e$, where we recall that $\Lambda_e$ is the number of extra bonds we add to $\Lambda$ to implement periodic b.c.

The precise definition of gauged away bonds, for free b.c., is as follows (see page 4 of \cite{bQCD}). We label the sites of the $\mu$-th lattice coordinate by $1,2,\ldots,L$. The enhanced temporal gauge is defined by setting in $\Lambda$ the following bond variables to $1$. First, for any $d =2,3,4$, we gauge away all temporal bond variables by setting $g_{b^0(x)}=1$. For $d=2$, take also $g_{b^1(x0=1,x1)}=1$. For d = 3, set also $g_{b^1(x0=1,x1,x2)}=1$ and $g_{b^2(x0=1,x1=1,x2)}=1$. Similarly, for $d=4$, set also to $1$ all $g_{b^1(x0=1,x1,x2,x3)}$, $g_{b^2(x0=1,x1=1,x2,x3)}$ and $g_{b^3(x0=1,x1=1,x2=1,x3)}$. For $d=2$ the gauged away bond variables form a comb with the teeth along the temporal direction, and the open end at the maximum
value of $x^0$. For $d=3$, the gauged away bonds can be visualized as forming a scrub brush with bristles along the $x^0$ direction and the grip forming a comb. For any $d$, all gauged away bond variables are associated with bonds in the hypercubic lattice $\Lambda$ which form a maximal tree. Hence, by adding any other bond to this set, we form a closed loop.

We now obtain factorized stability bounds for the partition function $Z^B_{\Lambda,a}$. In doing this, we are improving the proofs of \cite{YM,bQCD} and are extending the results to the periodic b.c. case. The bounds are factorized as a product. In the product, each factor is a single bond variable, single plaquette partition function. First, we give a Lemma that shows that the plaquette action $A_p$ has a global upper bound which is quadratic in each gluon bond variables.  The lemma is used to obtain the factorized lower bound on $Z^B_{\Lambda,a}$. The following Lemma is the content of Lemma 2 of Ref. \cite{bQCD}.
\begin{lemma}	\lb{lema1}
For the four retained bond plaquette, we have the global quadratic upper bound
\bequ\lb{lower1}
\mathcal A_p\,=\,\|U_p-1\|^2_{H-S}\,\leq\, C^2 \,\sum_{1\leq j\leq 4}\, |x^j|^2\,=\, C^2 \,\sum_{1\leq j\leq 4}\, |\la_j|^2\quad\:\:,\:\:\quad C=2\sqrt N\,,
\eequ
where $U_p\,=\,e^{iX_1}e^{iX_2}e^{iX_3}e^{iX_4}$. For $\alpha=1,\ldots,N^2$, $x^j_\alpha\,=\,\tr X_j\theta_\alpha$, and $\lambda_j=(\lambda_{j,1},\ldots,\lambda_{j,N})$, where, for $k=1,\ldots,N$, $\lambda_{j,k}$ are the angular eigenvalues of $e^{iX_j}$.

When there are only one, two or three retained bond variables in a plaquette, the sum over $j$ has, respectively, only one, two and three terms and the numerical factor $4$ in $C^2=4N$ is replaced by 1, 2 and 3, respectively. For the total action $A^B\,=\,\sum_p\,\mathcal A_p$, we have the global quadratic upper bound
\bequ\lb{lower2}A^B\,\leq\,2(d-1)\,C^2\,\sum_b|x^b|^2\,=\,2(d-1)\,C^2\,\sum_b|\la_b|^2\,,
\eequ
where the sum runs over all $\La_r$ retained bonds.
\end{lemma}

For completeness of the present paper, following Ref. \cite{bQCD}, we give the proof of Lemma \ref{lema1} in section \ref{sec4}. All the four theorems stated below are also proved there.

Our stability bounds on the partition function $Z^B_{\Lambda,a}$, leading to TUV stability bounds for the normalized partition function $Z^{B,n}_{\Lambda,a}$ are given by
\begin{thm} \lb{thm1}
The partition function $Z^B_{\Lambda,a}$ verifies the following stability bounds:  \vspace{2mm}\\
\noindent\underline{Free b.c.}:
\bequ\lb{sbfree}z_\ell^{\Lambda_r}\,\leq\,Z_{\Lambda,a}\,\leq z_u^{\Lambda_r}\,,\eequ
\noindent\underline{Periodic b.c.}:
\bequ\lb{sbperiodic}z_\ell^{\Lambda_r+\Lambda_e}\,\leq\,Z^P_{\Lambda,a}\,\leq\,Z_{\Lambda,a}\,\leq \,z_u^{\Lambda_r}\,,\eequ        
where
\bequ\lb{zu}
z_u\,=\,\int\, \exp\left[-2(a^{d-4}/g^2)\,\Re\tr (1-U)]\right]\; d\sigma(U)\,,
\eequ
and, with $U= e^{iX}$ and $C^2=4N$,
\bequ\lb{zl}
z_\ell\,=\,\,=\,\int\, \exp\left[-2C^2(a^{d-4}/g^2)\,(d-1)\,\tr X^2\right]\; d\sigma(U)\,.
\eequ
\end{thm}

\begin{rema}
Using Jensen's inequality, we obtain the factorized lower bound $Z_{\Lambda,a}\geq \xi^{\Lambda_p}$, where 
$$
\xi\,=\,\exp\left\{- \dfrac{a^{d-4}}{g^2}\,\dis\int\,\|U-1\|_{H-S}^2\,d\sigma(U) \right\}\,\geq\,\exp\left[- 2N\,\dfrac{a^{d-4}}{g^2} \right]\,,
$$
where we recall $\Lambda_p$ is the number of plaquettes in $\Lambda$. $\Lambda_p=\Lambda_r$, for $d=2$; $\Lambda_p\simeq 3L^3,6L^4$, respectively, for $d=3,4$. In Theorem \ref{thm2} below, we obtain factorized lower and upper bounds with $\Lambda_r=(d-1)L^d$ factors. In both the upper and lower bound a factor of $(a^{d-4}/g^2)^{-N^2/2}$ is extracted. This factor dominates the $a$, $g^2$ dependence.
\end{rema}

We continue by giving more detailed bounds for $z_u$ and $z_\ell$.  In these bounds, we extract a factor of $(a^{d-4}/g^2)^{-N^2/2}$ from both $z_u$ and $z_\ell$.  Note that the integrands of both $z_u$ and $z_\ell$ only depend on the angular eigenvalues of the gauge variable $U$; they are class functions on $\mathcal G$.  The $N^2$-dimensional integration over the group can be reduced to an $N$-dimensional integration over the angular eigenvalues of $U$ by the Weyl integration formula \cite{Weyl,Bump,Simon2}, which reads
\bequ\lb{weyl} \int_{\mathrm U(N)}\,f(U)\;d\sigma(U)\,=\,\dfrac1{\mathcal N_C(N)}\,\int_{(-\pi,\pi]^N}\,f(\lambda)\,\rho(\lambda)\,d^N\lambda\,,
\eequ
where $\mathcal N_C(N)=(2\pi)^N\,N!$, $\lambda=(\lambda_1,\ldots,\lambda_N)$, $d^N\lambda=d\lambda_1\ldots d\lambda_N$ and $\rho(\lambda)=\prod_{1\leq j<k\leq N}\,|e^{i\lambda_j}-e^{i\lambda_k}|^2$.

In our stability and generating functional bounds the following integrals of the Gaussian unitary ensemble(GUE) and Gaussian symplectic ensemble probability distributions (see \cite{Metha,Deift}), from random matrix theory, arise naturally.  Let, for $\beta=2,4$ and $u>0$,
$$I_\beta(u)\,=\,\dis\int_{(-u,u)^N}\,\exp\left[-(1/2)\,\beta\,\sum_{1\leq  j\leq N}\,y_j^2\right]\;\hat\rho^{\beta/2}(y)\, d^Ny\,,$$
where $\hat\rho(y)=\prod_{1\leq j<k\leq N}\,(y_j-y_k)^2$, $I_\beta(u)<I_\beta(\infty)=\mathcal N_\beta$, is the normalization constant for the GUE and the Gaussian Symplectic Ensemble (GSE) probability distributions for $\beta=2,4$, respectively. Explicitly, we have
$\mathcal N_G\,= (2\pi)^{N/2}\,2^{-N^2/2}\,\prod_{1\leq j\leq N}\,j!$ and  $\mathcal N_S\,= (2\pi)^{N/2}\,4^{-N^2}\,\prod_{1\leq j\leq N}\,(2j)!$.

For the upper bound on $z_u$ and lower bound on $z_\ell$, we have: 
\begin{thm} \lb{thm2}
Let $C^2=4N$. Then, we have the bounds    
\bequ\lb{uzu}\begin{array}{lll}
z_u&=& \mathcal N_C^{-1}\,\int_{(-\pi,\pi]^N}\,\exp[-2(a^{d-4}/g^2)\,\sum_{1\leq j\leq N}\,(1-\cos\lambda_j)]\;\rho(\lambda)\;d^N\sigma(\lambda)\vspace{1mm}\\
&\leq&(a^{d-4}/g^2)^{-N^2/2}\,(\pi/2)^{N^2}\,\mathcal N_G(N)\mathcal N^{-1}_C(N)\vspace{1mm}\\
&\equiv&(a^{d-4}/g^2)^{-N^2/2}\,e^{c_u}\,,\end{array}\eequ
and
\bequ\lb{lzl}\begin{array}{lll}
z_\ell&=& \mathcal N_C^{-1}\,\int_{(-\pi,\pi]^N}\,\exp[-2C^2(d-1)(a^{d-4}/g^2)\,\sum_{1\leq j\leq N}\,\lambda_j^2]\;\rho(\lambda)\;d^N\sigma(\lambda)\vspace{1mm}\\
&\geq&(a^{d-4}/g^2)^{-N^2/2}\,\mathcal N^{-1}_C(N)\,(4/\pi^2)^{N(N-1)/2}\,[2(d-1)C^2]^{-N^2/2}\,I_\ell\,,\vspace{1mm}\\
&\equiv&(a^{d-4}/g^2)^{-N^2/2}\,e^{c_\ell}\,,\end{array}\eequ
where  $I_\ell\equiv I_2(\pi[2(d-1)C^2]^{1/2}/(2g_0))$. $c_u$ and $c_\ell$ are finite real constants, independent of $a$, $a\in(0,1]$ and $g^2\in(0,g_0^2]$, $0<g_0<\infty$.
\end{thm}

Concerning the existence of the thermodynamic and continuum limits of the normalized free energy we define a normalized partititon function by
\bequ\lb{nZ}Z^{B,n}_{\Lambda,a}\,=\,(a^{d-4}/g^2)^{(N^2/2)\Lambda_r}\; Z^{B}_{\Lambda,a}\,,\eequ
and a finite lattice normalized free energy by
\bequ\lb{nf}f^{B,n}_{\Lambda,a}\,= \Lambda_r^{-1}\, \ln Z^{B,n}_{\Lambda,a}\,.\eequ
     
Using Theorem \ref{thm1} and Theorem \ref{thm2}, together with the Bolzano-Weierstrass theorem, we can directly prove the following Theorem.
\begin{thm}	\lb{thm3}
The normalized free energy $f^{B,n}_{\Lambda,a}$ converges subsequentially, at least, to a thermodynamic limit
$$f^{B,n}_{a}\,=\,\lim_{\Lambda\nearrow a\mathbb Z^d}\; f^{B,n}_{\Lambda,a}\,,$$ and, subsequently, again, at least subsequentially, to a continuum limit $f^{B,n}\,=\,\lim_{a\searrow 0}\,f^{B,n}_{a}$. Besides, $f^{B,n}_a$        satisfies the bounds
\bequ\lb{bdsf}
-\infty \,<\, c_\ell\,\leq\, f_a^{B,n}\,\leq c_u<\infty\,.
\eequ
and so does $f^{B,n}$. The constants $c_\ell$ and $c_u$ are finite real constants independent of $a\in(0,1]$ and $g^2\in(0,g^2_0]$, $0 < g_0< \infty$.
\end{thm}
\section{Generating Functional for Plaquette Field Correlations}\lb{sec3}
Here, we obtain bounds for the generating functional of gauge invariant field correlations. Bounds for the field correlations follow from Cauchy estimates on the generating functional. The same hypercubic lattice $\Lambda$ is maintained.  We use periodic boundary conditions and the multiple reflection method.  Our choice of correlations is guided by the energy-momentum spectral results from strong coupling (see \cite{Schor}). We fix $a=1$ and denote the plaquette coupling constant by $\gamma=a^{d-4}/g^2$.  For $0<\gamma\ll 1$, a lattice quantum field theory is constructed via a Feyman-Kac formula. By polymer expansion methods, infinite lattice correlations exist and are analytic in $\gamma\in\mathbb C$, $|\gamma|\ll 1$.  In \cite{Schor}, it is shown that, for $0<\gamma\ll 1$, associated with the truncated plaquette-plaquette correlation, there is an isolated particle (glueball) in the low-lying E-M spectrum, with mass of order $(–\ln \gamma)$.  Furthermore, for an arbitrary gauge-invariant function with finite support, it is shown that the isolated dispersion curve of the glueball is the only low-lying spectrum that is present.
Returning to our model, we consider the generating functional for the correlation of $r$ gauge-invariant real plaquette fields $\tr M_p(U_p)$.  Taking $p$ to be the plaquette $p_{\mu\nu}(x)$, $M_p(U_p)$ is defined by, with $U_p= e^{iX_p}$,
\bequ\lb{Mfield}\barr{lll}
M_p(U_p)&=&\dfrac{i}{2}\,\left(a^{d-4}/g^2\right)^{1/2}\,\left[(1-U_p)\,–\,(1 – U^\dagger_p)\right]\vspace{1mm}\\
&=&\left(a^{d-4}/g^2\right)^{1/2}\,\sin X_p\vspace{1mm}\\
&\simeq&-\,a^{(d-2)/2}\,\left\{\left[A_\nu(x+ae^\mu)\,–\,A_\nu(x)\right]\right\} \,–\,\left[A_\mu(x+ae^\nu)\, –\,A_\mu(x)\right]\,+\,iga[A_\mu(x),A_\nu(x)]\vspace{1mm}\\
&=&a^{d/2}\, F^a_{\mu\nu}(x)\,.\end{array}\eequ
Hence, 
$$\barr{lll}\tr M_p(U_p)&=&(a^{d-4}/g^2)^{1/2}\,\Im \tr (U_p -1)\,=\,       (a^{d-4}/g^2)^{1/2}\, \tr \sin X_p\vspace{1mm}\\
&\simeq&\,a^{d/2}\, \tr F^a_{\mu\nu}(x)\,.\end{array}$$

With our choice of the scaling factor $(a^{d-4}/g^2)^{1/2}$, the generating functional is finite, uniformly in $a\in(0,1]$. It may seem surprising that the generating functional is pointwise bounded.  However, it is known that a similar phenomenon occurs in the case of a free massless or massive real scalar field in $d=3,4$.  Namely, as analyzed in \cite{MP2}, if instead of the given physical field $\phi^u(x)$, we use a locally scaled field $\phi(x)\simeq a^{(d-2)/2}\phi^u(x)$, then the $r$–point correlation function for the scaled $\phi$ fields is bounded pointwise, uniformly in $a\in(0,1]$. No smearing by a smooth test function is needed to achieve boundedness! We give more detail in Appendix A.

Letting $p=p_{\mu\nu}(x)$, the $r$-plaquette generating functional is defined by
$$G_{r,\Lambda,a}(J^{(r)}) = \dfrac1{Z^P_{\Lambda,a}}\;Z^P_{r,\Lambda,a}(J^{(r)})\,,$$
where $Z^P_{r,\Lambda,a}(J^{(r)})$ is given by $Z_{\Lambda,a}^P$ with the inclusion of $r$ local source factors $\exp(\sum_{x\in\Lambda}\sum_{1\leq j\leq r}\,J_j(x_j)\tr M_{p_j}(U_{p_j}))$ and $J_j$,  $j = 1,\ldots,r$, are source strengths and with the convention that the plaquette $p_j$ originates at the lattice point $x_j$. The $r$-plaquette correlation, with a set $y_E\,=\,(y_1,\dots,y_r)$ of $r$ lattice external points in $\Lambda$ is given by  
$$\left.\dfrac{\partial^r}{\partial J_1(y_1)\ldots\partial J_r(y_r)}\,G_{r,\Lambda,a}(J^{(r)})\right|_{J_j=0}\,.$$

Our factorized bound is given in the next Theorem. For simplicity of notation, from now on, we set $J_i\,\equiv\,J_i(y_i)$.
\begin{thm} \label{thm4}
Considering the model with periodic b.c., we have:
\begin{enumerate}
\item The plaquette field generating functional is bounded by
\bequ\lb{Gb}|G_{r,\Lambda,a}(J^{{(r)}})|\,\leq\,\prod_{1\leq j\leq r}\,\dfrac{\left|z_u(rJ_j)\right|^{2^d\Lambda_r/(r\Lambda_s)}}{z_\ell^{2^d(\Lambda_r+\Lambda_e)/(r\Lambda_s)}}\,.\eequ  
\item From this, if $G_{r,a}(J^{(r)})$ denotes a sequential or subsequential thermodynamic limit $\Lambda\rightarrow a\mathbb Z^d$, then 
$$\left|G_{r,a}(J^{(r)})\right|\,\leq\, \prod_{1\leq j\leq r}\,\left|z_u(rJ_j)/z_\ell\right|^{2^d(d-1)/r}\,,$$
with
\bequ\lb{Gb2}\begin{array}{lll}
\left|z_u(J)\right|&=&\dis\int\,\exp\left[\,|J|\,(a^{d-4}/g^2)^{1/2}\,|\Im Tr(U-1)|\,-\,2(a^{d-4}/g^2)\,A_p(U)\,\right]\, d\sigma(U)\,\vspace{1mm}\\
&=& (\mathcal N_c)^{-1}\,\dis\int\,exp\left[|J|\,(a^{d-4}/g^2)^{1/2}\,\sum_{1\leq j\leq N}\, |\sin \lambda_j|\,-\,2a^{d-4}/g^2 \sum_{1\leq j\leq N}\,(1 – \cos \lambda_j )\right]\,\rho(\lambda)\,d^N\lambda\vspace{2mm}\\
&\leq&\dfrac{(a^{d-4}/g^2)^{-N^2/2}\,\pi^{N^2+N/4}\,\mathcal N_S^{1/2}}{\mathcal N_C}\:\exp(\pi^2/8N|J|^2)\vspace{2mm}\\
&=&(a^{d-4}/g^2)^{-N^2/2}\,\exp(c_u^\prime+\pi^2/8N|J|^2)\,.\end{array}\eequ
Hence, from the bounds of Eqs.(\ref{Gb}) and (\ref{Gb2}), it follows that $G_{r,\Lambda,a}(J^{{(r)}})$ is a jointly analytic, entire complex function of the source field strengths $J_j\in\mathbb C$.  
\item Letting $G_{r}(J^{(r)})$ denote a sequential or subsequential continuum limit $a\searrow 0$ of $G_{r,a}(J^{(r)})$, then 
$$\left|\,G_{r}(J^{(r)})\,\right|\,\leq\, \exp\left[2^d(d-1)\,(c^\prime-c_\ell)\,+\,(\pi^2/8)\,\sum_{1\leq j\leq r}\,    |J_j|^2\right]\,.$$
\end{enumerate}
This bound is independent of the location and orientation of the $r$ plaquettes, and independent of the value of $a\in(0,1]$ and $g^2$.
\end{thm}
\begin{rema}
By Cauchy estimates, the $r$-plaquette scaled field correlations are bounded. In particular, the coincident point plaquette-plaquette physical field correlation is bounded by $\mathrm{const}\,a^{2-d}$. The $a^{2-d}$ factor is the same small $a$ behavior we have for the coincident two- point correlation of the real scalar free field (see Appendix A). This is a signal of the UV asymptotic freedom.
\end{rema}

Lemma \ref{lema1} and Theorems $1-4$ are proved in the next section.
\section{Proofs of the Lemma and Theorems}\lb{sec4}
Here, following Ref. \cite{bQCD}, we give a proof of Lemma \ref{lema1}. We also prove Theorems $1-4$.
\subsection{Proof of Lemma \ref{lema1}}
For simplicity, here we consider the case where we have four retained bonds in a plaquette. The other cases are similar. We define, for $1\leq j\leq 4$, $\mathcal L_j=i\sum_{1\leq \alpha\leq N^2} x^j_\alpha\theta_\alpha$, so that  $U_j=e^{\mathcal L_j}$ and $U_p=U_1U_2U_3^\dagger U^\dagger_4$.

Since $\|\mathcal L_j\|\,\leq\,\|\mathcal L_j\|_{H-S}\,=\,|x^j|$ and letting $U_p(\delta)\,=\,U_1(\delta)U_2(\delta)U^\dagger_3(\delta)U^\dagger_4(\delta)$, $U_j(\delta)\,=\,e^{\delta\mathcal L_j}$, for $\delta\in[0,1]$, by the fundamental theorem of calculus, suppressing $\delta$,
$$
U_p\,-\,1\,=\,\dis\int_0^1\,d\delta\,\left[\mathcal L_1U_1U_2U_3^\dagger U_4^\dagger\,+\,U_1\mathcal L_2U_2U_3^\dagger U_4^\dagger\,-\,U_1U_2\mathcal L_3U_3^\dagger U_4^\dagger\,-\,U_1U_2U_3^\dagger\mathcal L_4 U_4^\dagger\right]\,.
$$
Using the triangle and Cauchy-Schwarz inequalities, we obtain
$$
\|U_p\,-\,1\|\,\leq\,\sum_{j=1}^4\,\|\mathcal L_j\|\,\leq\,\sum_{j=1}^4\,\|\mathcal L_j\|_{H-S}\,=\,\sum_{j=1}^4\,|x^j|\,\leq\,2\,\left[\sum_{j=1}^4\,|x^j|^2\right]^{1/2}\,.
$$
But, $\|U_p\,-\,1\|\,\geq\,N^{-1/2}\,\|U_p-1\|_{H-S}$. Hence,
$$
\mathcal A_p\,=\,\|U_p-1\|^2_{H-S}\,\leq\,4N\,\sum_{j=1}^4\,|x^j|^2\,.
$$
By considering the number of terms of the sum over $j$, the factor $4$ in $C^2$ is replaced by $1$, $2$ and $3$, respectively, when only one, two or three retained bond variables appear in a retained plaquette.

Using this upper bound on the single plaqquette action, and summing over the retained plaquettes, the second inequality of Lemma \ref{lema1} is easily proven.\qed
\subsection{Proof of Theorem \ref{thm1}}
\noindent{\em The Case of Free b.c.}:\vspace{1mm}\\
\noindent\underline{Upper Bound}:  Fix the enhanced temporal gauge.  An upper bound is obtained by discarding all horizontal plaquettes except those with temporal coordinates $x^0=1$.  We now perform the horizontal bond integration. For ease of visualization we carry it out explicitly for $d=3$. Integrate over successive planes of bonds starting at $x^0=L$ and ending at $x^0=1$. For each horizontal bond variable integration, the bond variable appears in only one vertical plaquette.  After the integration, in principle, the integral still depends on the other bond variables of the plaquette.   However, using the left or right invariance of the Haar measure, the integral is independent of the other variables. In this way, we extract a factor $z_u$. In the total procedure, we integrate over the $\Lambda_r$   horizontal bonds, so that we extract a factor $z_u^{\Lambda_r}$.\vspace{2mm}\\
\noindent\underline{Lower Bound}:  Fixing the enhanced temporal gauge and using Lemma 1 gives the factorization and $z_\ell$.\vspace{4mm}

\noindent{\em The Case of Periodic b.c.}:\vspace{1mm}\\
\noindent\underline{Upper Bound}: Considering the positivity of each term in the model action of Eq. (\ref{action}), since $A^P\geq A$, we have
$$Z^P_{\Lambda,a}\,\leq\,\int\,e^{-A}\,dg^B\,=\,\int\, e^{-A}\,dg\,=\,Z_{\Lambda,a}\,\leq\,z_u^{\Lambda_r}\,.$$ 
\vspace{2mm}\\
\underline{Lower Bound}: Fix the enhanced temporal gauge. Use the global quadratic upper bound of Lemma \ref{lema1} on all $\Lambda_r\cup\Lambda_e$ bond variables. Thus, we have
$$Z^P_{\Lambda,a}\,\geq\,z_\ell^{\Lambda_r+\Lambda_e}\,,$$
where $U = exp(iX)$, $X =\sum_\alpha x_\alpha\theta_\alpha$.\qed
\subsection{Proof of Theorem \ref{thm2}}
The first line of $z_u$ is the application of Weyl`s integration formula \cite{Weyl,Bump,Simon2}. Use the inequality (see \cite{Simon3}) $(1-\cos x) \geq 2x^2/\pi^2$, $x \in [-\pi,\pi]$, in the action. and the inequality $(1-\cos x)\leq x^2/2$ in each factor of $\rho(\lambda)$. After making the change of variables $y \,=\,2[a^{(d-4)/2}/(\pi g)]\,\lambda$ and using the monotonicity of the integral, the result follows. For $z_\ell$, apply Weyl`s integration formula and use the inequality $2[1-\cos(\lambda_j-\lambda_k)]\,\geq\,(4/\pi^2)\, (\lambda_j-\lambda_k)^2$, $|\lambda_\ell|<\pi/2$ in each factor of the density $\rho(\lambda)$. Then, use the positivity of the integrand and restrict the domain of integration to $(-\pi/2,\pi/2]^N$. In making the change of variables $y\,=\, [a^{(d-4)/2}/g]\,C\,\sqrt{2(d-1)}\,\lambda$, the integral $I_2([a^{(d-4)/2}/g]\,C\,\sqrt{2(d-1)})\pi/2)$ appears. Since $I_2(u)$ is monotone increasing the integral assumes its smallest value for $a = 1$ and $g^2=g_0^2$.  
\qed
\subsection{Proof of Theorem \ref{thm3}}
For periodic boundary conditions and the lower bound, using Theorem \ref{thm1}, we have the finite volume lattice normalized free energy
$$\barr{lll}
f^{P,n}_{\Lambda,a}&=&\dfrac1{\Lambda_r}\,\ln Z^{P,n}_{\Lambda,a}\,=\,\dfrac1{\Lambda_r}\,\ln\left[ \dfrac{a^{d-4}}{g^2}\right]^{N^2\Lambda_r/2}\,+\,\dfrac1{\Lambda_r}\,\ln Z^{P}_{\Lambda,a}\vspace{2mm}\\
&\geq&\dfrac1{\Lambda_r}\,\ln\left[ \dfrac{a^{d-4}}{g^2}\right]^{N^2\Lambda_r/2}\,\,+\,\dfrac1{\Lambda_r}\,\ln z_\ell^{\Lambda_r+\Lambda_\ell}\,.\earr
$$

Continuing the inequality and using Theorem \ref{thm2}, we have
$$\barr{lll}
f^{P,n}_{\Lambda,a}&\geq&\dfrac1{\Lambda_r}\,\ln\left[ \dfrac{a^{d-4}}{g^2}\right]^{N^2\Lambda_r/2}\,+\,\dfrac{\Lambda_e+\Lambda_r}{\Lambda_r}\,\ln \left[\left(\dfrac{a^{d-4}}{g^2}\right)^{-N^2/2}\,e^{c_\ell}\right]\vspace{2mm}\\
&\geq&\ln\left[ \dfrac{a^{d-4}}{g^2}\right]^{N^2/2}\,\,+\,\dfrac{\Lambda_e+\Lambda_r}{\Lambda_r}\,\left[\ln \left(\dfrac{a^{d-4}}{g^2}\right)^{-N^2/2}\,+\,c_\ell \right]\,\earr
$$
which gives, when $\Lambda\rightarrow a\mathbb Z^d$,
$$
f^{P,n}_{a}\,\geq\,c_\ell\,.
$$

A similar calculation for the upper bound, with $\Lambda_e=0$, proves the theorem for the upper bound. For free b.c., set $\Lambda_e=0$ in the above calculations.\qed
\subsection{Proof of Theorem \ref{thm4}}
To prove Theorem 4, first use the  generalized Holder`s inequality to bound $G_{r,\Lambda, a}(J^{(r)})$ by a product of single plaquette generating functionals, i.e.
$$|G_{r,\Lambda,a}(J^{(r)})|\,\leq\,\prod_{1\leq j\leq r}\,|G_{1,\Lambda, a}(rJ_j)|^{1/r}\,
$$
Now, since we are adopting periodic b.c., we can apply the multi-reflection method (see \cite{GJ}) to bound each factor in the product. To this end, we make a shift in the lattice by $(1/2a)$ in each coordinate direction. Also, we use the $\pi/2$ lattice rotational symmetry and translational symmetry to put the single plaquette in the $\mu\nu=01$ plane in the first quadrant with lower left vertex at $(a/2,a/2,\ldots,a/2)$. Then, we apply the multi-reflection method to obtain the bound 
$$|G_{1,\Lambda, a}(rJ_j)|\,\leq\,|G_{\Lambda,a}(rJ_j)|^{2^d/\Lambda_s}\,,$$
where $G_{\Lambda,a}(J)\,=\,\left[Z^P_{\Lambda,a}\right]^{-1}\,Z^P_{\Lambda,a}(J)$, with $Z^P_{\Lambda,a}(J)$ denoting $Z^P_{\Lambda,a}$ with a source of uniform source strength $J$.  The source factor is given by $exp[J\sum^\prime_p\tr M_p(U_p)]$, where the sum is over an array of plaquettes. The array consists of planes of plaquettes that are parallel to the $01$ plane. In each plane, there are only alternating, i.e. like considering only squares of a same color on a chessboard.  We obtain a greater upper bound by noting that
$$\barr{lll}
|J \tr M_p(U_p) |&\leq& |J|\,[a^{(d-4)/2}/g]\,|\Im \tr(U_p-1)|\\
&\leq&|J|\,[a^{(d-4)/2}/g]\,|\tr(U_p-1)|\\
&\leq&|J|\,[a^{(d-4)/2}/g]\,N^{1/2}\,\|U_p-1\|_{H-S}\,,\earr
$$
where we have used the Cauchy-Schwarz inequality in the Hilbert-Schmidt inner product.

We also increase the bound by summing over all plaquettes in the lattice $\Lambda$ that are parallel to the $01$ plane. We denote this sum by $\sum_p^{\prime\prime}$. In this way, we obtain the upper bound
$$
\left|Z^P_{\Lambda,a}(J)\right|\,\leq\,\int\, exp\left[|J| a^{(d-4)/2}g^{-1}N^{1/2}\sum^{\prime\prime}_p\,\|U_p-1\|_{H-S}\,-\,a^{d-4}A^P/g^2\right]\,dg^P\,.
$$

As in the proof of the upper stability bound for the periodic model, as given above, we discard plaquette actions in  $A^P$ for plaquettes that are not in $\Lambda$ so that 
$$
\left|Z^P_{\Lambda,a}(J)\right|\,\leq\,\int\, exp\left[|J| a^{(d-4)/2}g^{-1}N^{1/2}\sum^{\prime\prime}_p\,\|U_p-1\|_{H-S}\,-\,a^{d-4}A/g^2\right]\,dg\,.
$$

We bound the integral as we did for the upper stability bound for the free b.c. case.  In this manner, we obtain the factorized bound
$$
\left|Z^P_{\Lambda,a}(J)\right|\,\leq\,[z_u(J)]^{\Lambda_r}\,,
$$
and the factorized bound of the theorem for $G_{r\Lambda a}(J^{(r)})$ is proved.

Now, recalling that $\Lambda_s=L^d$, $\Lambda_r\simeq (d–1) L^d$ and $\Lambda_e=dL^{d-1}$, the factorized bound for $G_{ra}(J^{(r)})$ follows.

Application of the Weyl integration formula \cite{Weyl,Bump,Simon2} gives the  $\lambda$ integral for $z_u(J)$. Using the bounds $|\sin \lambda_j|\leq |\lambda_j|$, for all $j$, and  $|\exp(i\lambda_j)-\exp(i\lambda_k)|^2= 2[1 –\cos(\lambda_j-\lambda_k)]\leq (\lambda_j-\lambda_k)^2$, for each factor of $\rho(\lambda)$ gives the inequality
$$
z_u(J)\,\leq\,(1/\mathcal N_c)\,\displaystyle\int_{(-\pi,\pi]^N)}\,\exp\left[|J| (a^{(d-4)/2}/g)\,\sum_{1\leq j\leq N}|\lambda_j|\,-\,4a^{d-4}/(g^2\pi^2)\,\sum_{1\leq j\leq N}\,\lambda_j^2\right]\,\hat\rho(\lambda)\,d^N\lambda\,.
$$

Making the change of variables $y_k=[2a^{(d-4)/2}/(g\pi)]\,\lambda_k$, a factor of $[a^{(d-4)/2}/g]^{-N^2}$ is extracted and the remaining integral is bounded by, with $y^2=\sum_j\,y_j^2$,
$$
\dis\int_{\mathbb R^N}\,exp\left[\pi |J|\sum_{1\leq j\leq N} |y_j|/2\,-\,\sum_{1\leq j\leq N}y_j^2\right]\,\hat\rho(y)\,d^Ny\,.
$$

Writing $\exp(-y^2)=\exp(-y^2/2) \exp(-y^2/2)$ and using the Cauchy-Schwarz inequality, the integral is bounded by
$$
\left[\dis\int_{\mathbb R^N}\,exp\left(\pi\,|J|\sum_{1\leq j\leq N} |y_j|\,-\,\sum_{1\leq j\leq N}y_j^2\right)\,d^Ny\right]^{1/2}\;\left[ \dis\int_{\mathbb R^N}\,exp\left(-\,\sum_{1\leq j\leq N}y_j^2\right)\,\hat\rho^2(y)\,d^Ny\right]^{1/2}\,.
$$
Using the inequality $e^{s|y_k|}\leq e^{sy_k} + e^{-sy_k}$, $s>0$, the Gaussian integral of the bound of the integral of the first factor is carried out explicitly.  For the integral of the second factor, after making the change of variables $w_k=(y_k/\sqrt{2})$ and up to a numerical factor, the resulting integral is the normalization constant  $\mathcal N_S$ for the Gaussian symplectic ensemble \cite{Metha}. Keeping track of the numerical factors gives the final inequality for $z_u(J)$ and the proof of Theorem \ref{thm4} is complete.\qed      
\section{Concluding Remarks}\lb{sec5}
We consider the Yang-Mills relativistic quantum field theory in an imaginary-time functional integral formulation. In the spirit of the lattice approximation to the continuum, the Wilson partition function is used as an ultraviolet regularization, where the hypercubic lattice $\Lambda\subset a\mathbb Z^d$, $d=2,3,4$, $a\in (0,1]$, has $L$ sites on a side. We use both free and periodic boundary conditions and our lattice has $\Lambda_s=L^d$ sites.

If $x=(x^0,\ldots,x^{d-1})$ denotes a site of $\Lambda$ and  $e^\mu$, $\mu=0,\ldots,(d-1)$ is a unit vector in the positive $\mu$ direction ($0$ is the label of the time direction), the partition function for free and periodic boundary conditions is given by
$Z^B_{\Lambda,a}\,=\, \int\; \exp[(-a^{d-4}/g^2)\,A^B]\, d\tilde g^B$,
where $B=P$, for periodic b.c. and for free b.c. we omit the superscript. For each bond $b$, we assign a gauge bond variable $U_b\in\mathcal G$, where $\mathcal G$ is the gauge group $\mathrm U(N)$.  We denote by $b_\mu(x)$ the bond with the lattice initial point $x$ and terminal point $x+ae^\mu$.

Parametrizing the bond variable $U_b$, $b\equiv b_\mu(x)$, by $\exp[iagA_\mu(x)]$, we call the self-adjoint gauge potential $A_\mu(x)$ the physical gluon field. A lattice plaquette (minimal square) with vertices $x$, $x+ae^\mu$, $x+ae^\mu+ae^\nu$, $x+ae^\nu$, $\mu<\nu$, is denoted by $p_{\mu\nu}(x)$ and the model action $A^B$ is a sum over all plaquettes of four bond variable single plaquette actions $A_p$ of each plaquette $p_{\mu\nu}(x)$. Defining $U_p\,=\, e^{iagA_\mu(x)}\,e^{iagA_\nu(x+ae_\mu)}\,e^{-iagA_\mu(x+ae_nu)}\,e^{-iagA_\nu(x)}$, the plaquette action $A_p$ for the plaquette $p$ is givn by $A_p\,=\,\|U_p-1\|^2_{H-S}\,=\,2\,\Re\tr(1-U_p)\,=\,2\tr(1–\cos X_p)$, where we used the ordinary Hilbert-Schmidt norm and $U_p= e^{iX_p}$. $A_p$ is pointwise nonnegative and so is $A^B=\sum_p\,A_p$.

With this, the gauge group measure $d\tilde g^B$ above is a product over single bond $\mathcal G$ Haar measures $d\sigma(U_b)$ and, whenever periodic b.c. is employed, as usual, we add extra bonds to $\Lambda$ connecting the {\em endpoints} of the boundary to the {\em initial} points of the boundary of $\Lambda$ in each spacetime direction $\mu=0,1,\ldots,(d-1)$. The periodic plaquettes are those that can be formed from the totality of periodic bonds.

Formally, for small lattice spacing $a\in(0,1]$,
$A_p\,\simeq\,a^4g^2\;\tr [F^a_{\mu\nu}(x)]^2$
where, with finite difference derivatives understood, we have
$F^a_{\mu\nu}(x) = \partial^a_\mu A_\nu(x)\,–\,\partial^a_\nu A_\mu(x)\,+\,ig \,[A_\mu(x),A_\nu(x)]$ where the commutator is taken over the Lie algebra of $\mathcal G=\mathrm U(N)$. 
Thus, $(a^{d-4}/g^2)\,\sum_p A_p\,\simeq\,a^d\,\sum_{x\in\Lambda}\; \sum_{\mu,\nu=0,1,\ldots,(d-1)\,;\,\mu<\nu}\,\tr [F^a_{\mu,\nu}(x)]^2$
is the Riemann sum approximation to the classical smooth field continuum YM action
$\sum_{\mu<\nu}\,\int\,\tr [F_{\mu\nu}(x)]^2\,d^dx$, where $F_{\mu\nu}$ is defined as above but with usual partial derivatives.

Associated with this classical statistical mechanical model partition function and its correlations there is a lattice quantum field theory.  The quantum field theory is constructed in \cite{Sei} via a Feyman–Kac formula. An important ingredient in the construction is Osterwalder-Seiler reflection positivity which requires the number of lattice points $L$, in each spacetime diretion, to be even. The construction provides a quantum mechanical Hilbert space, mutually commuting self-adjoint spatial momentum operators and a positive energy operator.

Here, we define a normalized partition function $Z^{B,n}_{\Lambda,a}$, related to $Z^{B}_{\Lambda,a}$ by a $g$ and $a$-dependent multiplicative factor, and show that $Z^{B,n}_{\Lambda,a}$ obeys thermodynamic and ultraviolet stability bounds. These bounds guarantee the existence of a normalized free energy for a sequential or subsequential thermodynamic limit ($\Lambda\nearrow a\mathbb Z^d$) and, subsequently,  a subsequential continuum limit($a\searrow 0$). The proof given here has some improvements on the results of \cite{YM,bQCD} and also extends the results to the case when periodic boundary conditions are employed. The use of periodic conditions allows us to employ the multireflection method \cite{GJ} to prove  bounds for the plaquette fields generating functional which we also analyze here.

As a key ingredient for the lower bound on $Z^{B,n}_{\Lambda,a}$, we have found a new upper bound for the four-bond variable Wilson plaquette action. The bound is local and quadratic in the gluon bond variables of the plaquettes. It is surprising since the naive small $a$ approximation to the action has positive quartic terms in the case of a nonabelian gauge group.

The bounds have a product structure. The number of factors is roughly the number of bond variables in the temporal gauge, i.e. $(d-1)L^d$.

As before, for the case of free b.c., considering here also the use of periodic b.c., the upper (lower) stability bound factor is denoted by $z_u$ ($z_\ell$) and is a single bond variable, single-plaquette partition function.  We prove that a factor $\xi\equiv(a^{d-4}/g^2)^{-N^2/2}$ can be extracted from both $z_u$ and $z_\ell$ so that $z_u<\xi e^{c_u}$ and $z_\ell>\xi e^{c_\ell}$, with finite constants $c_\ell$ and $c_u$, independent of $a\in(0,1]$ and $g\in[0,g_0]$, $0<g_0<\infty$.

Using periodic b.c., we also obtain bounds for the normalized generating functional for $r\in\mathbb N$ plaquette fields defined, with a collection of $r$ source plaquette fields with source strengths $J^{(r)}=\{J_1,\ldots,J_r\}$, in the finite lattice $\Lambda$,  by
$$G_{r,\Lambda,a}(J^{(r)})\,=\,\dfrac{ Z^P_{r,\Lambda,a}(J^{(r)}) }
{Z^P_{\Lambda,a}}\,,$$
where $Z^P_{r,\Lambda,a}$ is the partition function $Z^P_{\Lambda,a}$ with the inclusion of the usual exponential of the source factors, namely,
$\exp[\sum_{1\leq j\leq r}\,J_j\, \tr M_{p_j}(U_{p_j})]$.
Here $p_j$, $j = 1,\ldots,r$ are plaquettes $p_{\mu_j\nu_j}(x_j)$. For fixed $\mu$ and $\nu$, and the plaquette $p_{\mu\nu}(x)$, the plaquette field we consider is approximately, for small $a$,
$$\tr M_p(U_p)\,=\,a^{d/2}\,\tr F^a_{\mu\nu}(x)\,.$$
(Note that the trace does not give zero since $\mu$ and $\nu$ are fixed!)
The $r$-plaquette field correlation, with plaquettes originating at the external points $x_E=\{x_1,\ldots,x_r\}$, is given by 
$$\mathcal G_{r,\Lambda,a}(x_E)\,=\left.\,[\delta/\delta (J_1(x_1)]\,\ldots\, [\delta/\delta (J_r(x_r)]\;G_{r,\Lambda,a}(x_E)\;\right|_{J_1,\ldots,J_r=0}\,.$$

We also prove a factorized bound for $\mathcal G_{r,\Lambda,a}(x_E)$ so that, denoting by $G_{ra}(J^{(r)})$ any sequential or subsequential thermodynamic limit, we have
$$G_{ra}(J^{(r)})\,\leq\,\prod_{j=1}^r \left[ \dfrac{z_u(J_j)}{z_\ell} \right]^{2^d(d-1)/r}\,,$$
where $z_u(J_j)$ is a single bond variable single plaquette partition function with a source of strength $J_j$.  It is shown that 
$$z_u(J)\,\leq\, (a^{d-4}/g^2)^{-N^2/2}\,e^{c^\prime_u}\,e^{cJ^2}\,,$$
where $c^\prime_u$ and $c$ are finite real constants, independent of $a$ and $g^2$. Thus,
$$G_{ra}(J^{(r)})\,\leq\,\exp\left[ 2^d(d-1)\,(c^\prime_u-c_\ell)\,+\,cr\sum_{1\leq j\leq r}\,J_j^2
\right]\,.$$

The bounds extend to complex source strengths. The generating functional is a jointly analytic, entire function in the source strengths $J_1$, ..., $J_r$ of the $r$ plaquette fields. The $r$-plaquette field correlations admit a Cauchy integral representation and are bounded by Cauchy bounds.

Our stability and generating functional bound results extend to the gauge group $\mathcal G-\mathrm{SU}(N)$ and other connected, compact Lie groups $\mathcal G$.  By the Bolzano-Weierstrass theorem, the stability bounds ensure the existence of a normalized free energy for the model, but do not give information on any other model property and its the energy-momentum spectrum. The existence of the normalized free energy and boundedness of the generating functional are the only questions we analyze here. More analysis is indeed needed e.g for the correlation decay rates. We point out that our bounds hold whether or not a mass gap persists in the $a\searrow 0$  continuum limit.         
\appendix{\begin{center}{\bf APPENDIX A: Unscaled or Physical and Scaled real scalar free fields}\end{center}}
\lb{appA}
\setcounter{equation}{0}
\setcounter{lemma}{0}
\setcounter{thm}{0}
\renewcommand{\theequation}{A\arabic{equation}}
\renewcommand{\thethm}{A\arabic{thm}}
\renewcommand{\thelemma}{A\arabic{lemma}$\:$}
\renewcommand{\therema}{{\em{\bf A\arabic{rema}$\:$}}}\vspace{.5cm}
Here, considering the case of the real scalar free field $\phi$, we develop the relation between quantities expressed in terms of the local unscaled or physical field $\phi^u(x)$ and locally scaled fields $\phi(x)\,=\,s\,\phi^u(x)$, with 
$$s\equiv s(a)=[a^{d-2}(m_u^2a^2+2d\kappa_u^2)]^{1/2}\,,$$
where $m_u$ and $\kappa_u$ are the unscaled field mass and lattice hopping parameter defined below. We refer to Ref. \cite{bQCD} for more details.

In the continuum limit, the unscaled two-point correlation is infinite at coincident points.  By our choice of $s$, for $d = 3,4$, the scaled field correlations are more regular in the continuum limit. More precisely, they are finite at coincident points! For the massless free scalar field, the $a$-dependence of the scaling factor is $a^{(d-2)/2}$.

In the case of YM, as discussed above, this same scaling factor relation between the physical gluon fields $A_\mu(x)$ and the scaled gluon fields $a^{(d-2)/2}\,A_\mu(x)$ makes the scaled plaquette correlations bounded, in the continuum limit.

Of course, these scaling transformations are not to be confused with the usual canonical scaling.

In the hypercubic lattice with free boundary conditions, the unscaled or physical action for the real scalar free field is, up to boundary conditions and for $x^+_\mu\equiv x+ae^\mu$,
\bequ\lb{sM}
\barr{lll}
A^u_{B,a}&=&\frac{\kappa_u^2}2\,a^{d-2}\,\sum_{x,\mu}\,\left[\phi^u(x^+_\mu)-\phi^u(x)\right]^2\,+\,\frac12\,m_u^2\,a^d\,\sum_{x}\,\left[\phi^u(x)\right]^2\vspace{2mm}\\
&=&-{\kappa_u^2}\,a^{d-2}\,\sum_{x,\mu}\,\phi^u(x^+_\mu)\phi^u(x)\,+\,\frac12\,(m_u^2\,a^d\,+\,2d\kappa_u^2a^{d-2})\,\sum_{x}\,\left[\phi^u(x)\right]^2\,.\earr
\eequ
$A^u_{B,a}$ is a sum of an unscaled hopping term, with an unscaled hopping parameter $\kappa_u^2>0$, and a mass term.

The thermodynamic limit of the unscaled two-point free field correlation exists and has the representation
$$
C^u_a(x,y)\,=\,\dfrac1{2(2\pi)^d}\,\int_{(-\pi/a,\pi/a]^d}\,e^{ip(x-y)
}\,D_a^{-1}\,d^dp\,.
$$
where $$D_a\,=\,\dfrac{\kappa_u^2}{a^2}\,\sum_\mu\,\left(1-\cos p_\mu a\right) \,+\, (m_u^2/2)\,.$$

The continuum limit $C^u(x,y)$ of $C_a^u(x,y)$ also exists, in the sense of distributions and is
$$C^u(x,y)\,=\,\dfrac1{(2\pi)^d}\,\int_{\mathbb R^d}\,\dfrac{e^{ip(x-y)}}{\kappa_u^2\sum_\mu\,(p^\mu)^2\,+\,m_u^2}\,d^dp\,,
$$
with $x,y\,\in\,\mathbb R^d$.
Of course,  $C^u(x,y)$ is infinite at coincident points $x=y$.

The formula for $C^u_a(x,y)$ is obtained as the thermodynamic limit of the finite lattice two-point correlation which in turn is obtained from the spectral representation of the symmetric matrix $M^{u,B}_{\Lambda,a}$ associated with the quadratic form, i.e. $S^{u,B}_{\Lambda,a}\,=\,\left(\phi^u,M^{u,B}_{\Lambda,a}\phi^u \right)$, with b.c. $B$. $B$ can be taken as free or periodic b.c.

The formula which relates the two-point correlation to the spectral representation is
$$
\barr{lll}
C^u_a(x,y)&=&\left[\,\dis\int\,\phi^u(x)\phi^u(y)\,e^{-S^{u,B}_{\Lambda,a}}\,d\tilde\phi^u\right]\:\left[\,\dis\int\,e^{-S^{u,B}_{\Lambda,a}}\,d\tilde\phi^u\right]^{-1}\,,\vspace{2mm}\\
&=&\dfrac12 \,\left[ M^{u,B}_{\Lambda,a}\right]^{-1}(x,y)\;=\;\dfrac12\,\sum_\upsilon\,(\lambda_\upsilon)^{-1}\,v_\upsilon^B(x)\,\left[v_\upsilon^B(y)\right]^t\,,\earr
$$
where $t$ denotes the transpose and we write the spectral representation of $M^{u,B}_{\Lambda,a}$ as
$$
M^{u,B}_{\Lambda,a}(x,y)\,=\,\sum_\upsilon\,\lambda_\upsilon\,v_\upsilon^B(x)\,\left[v_\upsilon^B(y)\right]^t\,,
$$
with $\lambda_\upsilon$ denoting an eigenvalue of  $M^{u,B}_{\Lambda,a}(\cdot,\cdot)$ and $v_\upsilon^B(\cdot)$ the corresponding eigenvector. The $\upsilon$'s, $\upsilon\,=\,(\upsilon^0,\ldots,\upsilon^{d-1})$, $\upsilon^\mu\in (-\pi/a,\pi/a]$, that parametrize the sum depend on the b.c. but the thermodynamic limit $C^u_a$ is the same for free and periodic b.c. For $m_u=0$, zero is (respectively, not) an eigenvalue of $M^{u,P}_{\Lambda,a}$ ($M^{u}_{\Lambda,a}$).

To obtain a more regular or less singular behavior for the correlations, as well as for the model free energy, we introduce the above defined locally scaled fields $\phi(x)$. With this scaling, unscaled field acion $A^u_{B,a}$ is transformed to the scaled action
\bequ\lb{sM2}
A_{B,a}({\tilde{\phi}})\,=\,-\,\kappa^2\,\sum_{x,\mu}\,\phi(x^+_\mu)\phi(x)\,+\,\frac12\,\sum_{x}\,[\phi(x)]^2\,,
\eequ
where the scaled hopping parameter to $\kappa^2$ is given by
\bequ\lb{hopbose}
\kappa^2\,=\,\left[2d\,+\,\left(\dfrac{m_ua}{\kappa_u}\right)^2\right]^{-1}\,.\eequ

The thermodynamic limit of the scaled 2-point correlation  is
$$
C_a(x,y)\,=\,\dfrac1{(2\pi)^d}\,\int_{(-\pi,\pi]^d}\,e^{ip(x-y)/a}\,\mathcal D^{-1}\,d^dp\,.
$$
with $\mathcal D\,=\,1\,-\,2\kappa^2\,\sum_\mu\,\cos q_\mu$. $C_a(x,y)$ is bounded uniformly in $a\in (0,1]$, for $d =3,4$, by the coincident-point value with $a=0$, namely, $C_0\,\equiv\,C(0)\,=\,(2\pi)^{-d}\,\int_{(-\pi,\pi]^d}\,[1 – d^{-1}\sum_\mu\,\cos q_\mu]^{-1}\,d^dq$, which is finite.

Thus, the scaled free field correlations are not singular, even at coincident points. Furthermore, the scaled and unscaled two-point correlations are related by
$C_a(x,y)\,=\,s^2\,C_a^u(x,y)$. Moreover, upon letting $C_a(x-y)\,\equiv\,C_a(x,y)$, the two-point correlation decay rate is defined by 
$\lim_{\upsilon\rightarrow\infty}\,(-1/\upsilon)\,\ln C_a(\upsilon)$, with $\upsilon\equiv x^0$. Thus, the decay rates are the same for $C^u_a(x)$ and $C_a(x)$.  Considering the Osterwalder-Schrader constructed lattice quantum field theory (see \cite{GJ,Sei}), the decay is the same as the scalar particle mass. The mass is a point in the energy-momentum spectrum with zero total spatial momentum. The mass $m$ satisfies $D_a(p^0=im,\vec p = 0)\,=\,(\kappa_u^2/a^2)\,(1 – \cosh ma) + m_u^2/2 = 0$, with solution
$$m\,=\,(2/a)\,\sinh^{-1}(m_ua/2k_u)\,=\,(m_u/\kappa_u)\,+\,\mathcal O(a^2(m_u/k_u)^3)\,.$$
It is important to observe that $m$ is jointly analytic in $a$ and $m_u$.
	
The above results continue to hold for the thermodynamic limit in the massless case $m_u=0$, for the case of free boundary conditions, as above.  For periodic boundary conditions, take the thermodynamic limit first with $m_u\not= 0$ and then take the limit $m_u=0$ to get the same result as for free boundary conditons.  The massless case is obtained by setting $\kappa^2\,=\,(1/2d)$ in the formula for $C_a(x,y)$.
In this case, the scaled field is related to the unscaled field by 
$\phi(x)\,=\,a^{(d-2)/2}\,\sqrt{2d}\,\kappa_u\,\phi^u(x)$, and we note that the $a$-dependence of the scaling factor is $a^{(d-2)/2}$.

The generating functional for the $r$-point scaled free field correlations is
$$
\exp\left[\dfrac 12\,\sum_{1\leq j,k\leq r}J_jC_a(x_j,x_k)J_k \right]\,\leq\,\exp\left[C_0r\sum_{1\leq j\leq r}\,J_j^2  \right]\,.
$$
The bound is uniform in $a\in (0,1]$, and is independent of the location of the $r$ points.

The generating function formula is obtained as the thermodynamic limit of the finite $\Lambda$ lattice generating functional. For the case of $r$ real variables $w_1,\ldots,w_r$, we use the conventional formula
$$
\left[\dis \int\,e^{(K,w)\,-\,(w,C^{-1}w)/2}\,d^rw \right]\:\left[\dis \int\,e^{-\,(w,C^{-1}w)/2}\,d^rw \right]^{-1}\;=\; e^{(K,CK)/2}\,,
$$
for the generating functional of $r$ source variables $w_1,\ldots,w_r$ with source strengths $K_1,\ldots,K_r$.

For the unscaled field $\phi^u$, with source factor $\exp[(\phi^u,f)]\,=\,\exp[\sum_ja^d\phi^u(x_j)f(x_j)]$, the generating functional is
$$
\exp\left[(1/2)\,\sum_{j,k}a^{2d}\,f(x_j)\,(\Delta^u)^{-1}(x_j,y_k)\,f(y_k)\right]\,,
$$
where
$$
(\Delta^u_a)^{-1}(x,y)\,=\,\dfrac1{2(2\pi)^d}\,\dis\int_{(-\pi/a,\pi/a]^d}\,e^{ip(x-y)}\,[D^u_a(p)]^{-1}\,d^dp\,.
$$
and
$$
D^u_a(p)\,=\,\left(\dfrac{\kappa_u}a \right)^2\,\sum_\mu\,(1-\cos p^\mu a)+m_u^2/2\,.
$$

$(\phi^u,f)$ is the Riemann sum approximation to $(\phi^u,f)_2\,=\,\int_{\mathbb R^d}\,\phi^u(x)f(x)\,d^dx$ and the $a\searrow 0$ limit of the generating functional is
$$
\exp\left[(1/2)\,\int_{\mathbb R^d}\,f(x)\,\Delta^{-1}(x,y)\,f(y)\right]\,d^dxd^dy\,=\,\exp\left[(1/2)\,\left( f,\Delta^{-1}f\right)_2\right]\,,
$$
where
$$\Delta^{-1}(x,y)\,=\,\int_{\mathbb R^d}\,e^{ip(x-y)}\,\dfrac 1{\kappa_u^2\sum_\mu (p^\mu)^2+m_u^2}\,d^dp\,.
$$
If $f=\Delta^{1/2}\,h$, $h\in L_2(\mathbb R^d,d^dx)$, the continuum generating functional is finite.
\begin{acknowledgements}
We would like to acknowledge the partial support of FAPESP and the Conselho Nacional de Desenvolvimento Cient\'\i fico e Tecnol\'ogico (CNPq).\vspace{3mm}
\end{acknowledgements}

\end{document}